\documentclass[prb,showpacs,floatfix,twocolumn,byrevtex]{revtex4}
\usepackage{graphicx}
\usepackage{dcolumn}
\usepackage{epsfig}
\usepackage{float}
\usepackage{bm}
\begin{document}

\bibliographystyle{apsrev}

\title{Infrared study of giant dielectric constant in Li and Ti doped NiO}
\author{Jung-Ho Kim$^{1}$, Youngwoo Lee$^{2}$, A Souchkov$^{3}$, J. S. Lee$^4$, H. D. Drew$^{3}$, S.-J. Oh$^{1}$, C.W. Nan$^{5}$ and E. J. Choi$^{2}$}

\affiliation{$^1$School of Physics Center for Strongly Correlated
Materials Research, Seoul National University, Seoul 151-742,
Republic of Korea} \affiliation{$^2$Department of Physics,
University of Seoul, Seoul 130-743, Republic of Korea}
\affiliation{$^4$Dept. of Physics, Univ. of Maryland, College
Park, MD20742, USA} \affiliation{$^4$School of Physics and
Research Center for Oxide Electronics, Seoul National University,
Seoul 151-747, Korea}\affiliation{$^5$Department of Materials
Science and Engineering, Tsinghua University, Beijing 100084,
China\\}

\begin{abstract}
We have measured optical reflectivity of Li and Ti doped NiO (LTNO) in the
infrared range at various temperatures. A Drude-like absorption is found at
low energy, $\omega <$ 100 cm$^{-1}$ and its spectral weight increases
substantially as temperature decreases. This observation and DC-resistivity
result show that LTNO has a conductive grain and resistive boundary. Such
composite structure provides evidence of the Maxwell-Wagner (MW) mechanism
as the origin of the high dielectric constant $\varepsilon _{o}$. We propose
a three-phase granular structure and show that this extended MW model
explains the observed frequency and temperature dependence of the dielectric
constant as well as the giant value of $\varepsilon _{o}$.
\end{abstract}

 \maketitle

The search for high dielectric constant ($\varepsilon _{o}$)
materials has been driven by the continuing demand for electronic
devices miniaturization. As a capacitor component, the high
dielectric constant enables the reduction of the circuit size and
can realize the tera-bit density static/dynamic random-access
memory.$\cite{kingon}$ It is also needed as a resonator and filter
in the microwave telecommunication. High $\varepsilon_{o}$ of
$\sim$ 1000 is found in ferroelectric oxides such as BaTiO$_{3}$
and in dielectric relaxors such as (Bi,Sr)TiO$_{3}$.\cite{chen}
However, in these materials, the dielectric constant is strongly
temperature dependent and the application is often limited.

The finding of huge dielectric constant in
CaCu$_{3}$Ti$_{4}$O$_{12}$(CCTO) has provided a new subject of
study.\cite{subramanian,ramirez} $\varepsilon_{o}$ of CCTO is high
($>10^5$) and has small frequency and temperature dependence which
are advantageous for practical use. Accumulating evidences exclude
this material from the ferroelectric/relaxor category. He
\textit{et al. }  made a first principle calculation of the
lattice and electronic structure of CCTO but found no direct link
of these intrinsic properties with the high
$\varepsilon_{o}$.\cite{he} As an alternative, extrinsic origin
associated with the composite microstructure has been
considered.\cite{subramanian,ramirez,kolev,he,sinclair} In the
internal barrier layer capacitor (IBLC) theory, the bulk CCTO is
assumed to consist of semi conductive domain and insulating
boundary, caused by the twinning of the unit cell.\cite{ramirez}
This composite capacitor structure can give rise to the high
$\varepsilon_{o}$ in the capacitance measurement through the
Maxwell-Wagner mechanism (MW). Also the decrease of
$\varepsilon_{o}$ when Ca is replaced by Cd was understood in this
picture. \cite{homes2}

Recently, Wu \textit{et al. } found that a Li and Ti co-doped NiO
(LTNO) sample exhibits a giant dielectric constant.\cite{wu} The
high $\varepsilon_{0}$ ($>10^4$) is maintained over wide frequency
range ($10^{2}$ Hz - $10^{6}$ Hz). As in the case of CCTO,
$\varepsilon_{0}$ drops above a characteristic frequency
$\Gamma(\sim 10^5$ Hz at 250 K). Even after the drop, the
dielectric constant is still significant, $\sim$300. This
frequency dependent behavior is described numerically by the Debye
form, $ \varepsilon (\omega )=\varepsilon _{\infty }+(\varepsilon
_{o}-\varepsilon _{\infty })/(1-i\omega /\Gamma )$ where
$\varepsilon _{o}$ ($\varepsilon _{\infty })$ represents the
static (high frequency) dielectric constant. $ \varepsilon _{o}$
is known to change with the Li and Ti concentration. As $T$
decreases, $\Gamma $ decreases drastically as $\Gamma(T)\sim
e^{-U/kT}.$ The high $\varepsilon _{o}$ is absent in NiO. Thus, a
comparison of LTNO and NiO through a spectroscopic study will
provide useful information on the origin of the high dielectric
constant.

In this work, we have made infrared reflectivity measurement of
LTNO and NiO to study the effect of Ni-site substitution on the
structural and electronic properties. The infrared features we
observe show that Li and Ti doping yields the IBLC structure in
LTNO. Then we analyze the high dielectric constant in terms of the
MW theory and find that various aspects of $ \varepsilon (\omega
)$ such as the $\omega$- and $T$-dependence can be coherently
explained.

The Li$_{0.05}$Ti$_{0.02}$Ni$_{0.93}$O (LTNO) and NiO pellets were
prepared by the sol-gel method as described in Ref. 9. Near-normal
incident reflectivity R$(\omega )$ was measured in the frequency
range of 25 - 5000 cm$^{-1}$ using a Fourier transform
spectrometor(Bomem DA8). The cubic structure of LTNO enables the
probe of intrinsic property with the polycrystal samples. The
sample was mounted in a closed cycle He-flow cryostat (Oxford) and
the temperature was varied between 50 and 300 K.

\begin{figure}[tbp]
\centering\epsfig{file=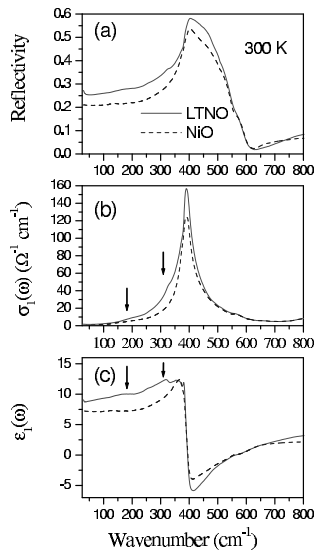,width=5cm}
\caption{Comparison of infrared spectra of LTNO and NiO: (a) Reflectivity
(b) Conductivity (c) Dielectric constant}
\label{fig:1}
\end{figure}

Figure 1 shows the infrared spectra of LTNO and NiO taken at room
temperature. In Fig.1(a), the reflectivity of NiO shows a typical
behavior of a dielectric insulator. R$(\omega)$ consists of a
IR-active phonon structure and a flat electronic background at low
frequency. The former is composed of a transverse optical (TO) and
a longitudinal optical (LO) mode of Ni-O oscillation which
corresponds to the peak at 400 cm$^{-1}$ and the dip at 620
cm$^{-1}$, respectively. In LTNO, the overall structure of R$
(\omega)$ is similar to that of NiO, except that the level is
somewhat higher.\cite{high}

From R$(\omega)$, we obtained the optical conductivity
$\sigma_{1}(\omega)$ (Fig.1(b)) and the dielectric constant
$\varepsilon_{1}(\omega)$ (Fig.1(c)) through the Kramers-Kronig
transformation. Here we used a constant-reflectivity extension for
the low frequency extrapolation $ \omega\longrightarrow0$. For the
high frequency side, R$(\omega)$ was extended to 20,000 cm$^{-1}$
above which the standard form R$ (\omega)\sim\omega^{-4}$ was
employed. In $\sigma_{1}(\omega)$, while the phonon spectra are
similar in the two samples, we find that two weak structures
appear at $\sim$ 320 cm$^{-1}$ and 200 cm$^{-1}$ in LTNO. These
absorptions may suggest a structural distortion due to the (Li,Ti)
substitution which can activate new phonon modes. In the $
\varepsilon_{1}(\omega)$ plot, these absorptions appear as
dispersive structures as indicated by the arrows. They contribute
to enhance $ \varepsilon_{1}$ but the strengths are weak and are
unlikely to be related with the high $\varepsilon_{o}$.

\begin{figure}[tbp]
\centering\epsfig{file=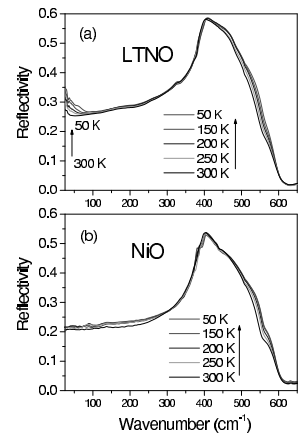,width=4cm,angle=0}
\caption{Temperature dependent reflectivity of (a) LTNO and (b) NiO}
\label{fig:2}
\end{figure}

Figure 2 displays R$(\omega )$ of LTNO and NiO at various
temperatures down to 50 K. In LTNO, we find two $T$-induced
changes as indicated by the arrows. As temperature decreases, the
reflectivity increases continuously in the phonon region around
550 cm$^{-1}$. The other feature occurs at low frequency below
$\sim $ 100 cm$^{-1}$ where a Drude-like absorption is found to
grow. The former increase occurs over the Restrahlen band,
$400-600$ cm$ ^{-1}$ where $\varepsilon _{1}(\omega )$ $<0$. The
reflectivity level in this range depends on the phonon damping
rate. At low $T$, the phonon damping is reduced and the Restrahlen
band becomes more reflecting. Except this minor change, the phonon
structure does not exhibit any $T$-dependent behavior in the
strength and frequency. This is in clear contrast with the case of
the ferroelectric materials such as SrTiO$_{3}$ and the cubic
perovskite CCTO where the phonon peaks change drastically due to
the lattice rearrangement or the charge redistribution within the
unit cell.\cite {homes2,homes} In LTNO, therefore, the lattice
effect is un- or minimally coupled with the high dielectric
constant. In NiO, note that the Drude-like feature is absent,
which shows that the feature is an effect of the Li / Ti doping.

Figure 3 shows this interesting low frequency absorption in an expanded
scale. The feature is weak at 300 K but grows continuously with decreasing $T
$ to a substantial amount. We think that this metallic feature is associated
with the Li (and possibly Ti) ions which can have different electronic
valency (+1 for Li) from that of Ni ion (+2). We used the Drude-Lorentz
model based on the classical dispersion theory to fit R$(\omega )$
\[
\varepsilon (\omega )=\varepsilon _{\infty
}+\sum_{i}\frac{S_{i}\omega _{i}^{2}}{\omega _{i}^{2}-\omega
^{2}-i\gamma _{i}\omega }+\frac{4\pi i}{ \omega }\sigma (\omega )
\]
Here $\varepsilon _{\infty }$ represents the high frequency
dielectric contribution. As for the phonon part (the second term),
we followed the analysis of Gielisse $et$ $al.$ who used two sets
of TO-LO modes to fit the phonon spectrum of NiO.\cite{gielisse}
S$_{i}$, $\omega _{i}$ and $\gamma _{i}$ represent the oscillator
strength, frequency and damping rate of the $i $-th mode. We also
included two weak absorptions at 200 and 320 cm$^{-1}$. In the
Drude term, $\sigma (\omega )=\omega _{p}^{2}/{4\pi (\gamma
-i\omega ) }$ represents the ac-conductivity where $\omega _{p}$
and $\gamma $ correspond to the plasma frequency and scattering
rate, respectively. The inset shows the fitting result over a wide
range. The fitting parameters are summarized in Table 1. As $T$
decreases, we find that $\omega _{p}$ and $ \gamma $ increase.
This behavior will be discussed later on. DC resistivity
calculated from $\rho =59.9\times \gamma /\omega _{p}^{2}$
$(\Omega \cdot cm) $ shows a moderate decrease with lowering $T$,
suggesting a poor metallic character of the carriers. In Eq.(1),
when $\omega $ is low enough than the phonon frequencies, the
first two terms can be simplified as $\varepsilon _{\infty
}+\sum_{i}S_{i}$ $\equiv $ $\varepsilon _{ir}$ and Eq.(1) is
reduced to $\varepsilon (\omega )=\varepsilon _{ir}+4\pi i\sigma
(\omega )/\omega $. We will use this convenient form when we
discuss below the high $ \varepsilon $ of the radio frequency
range, $\omega <10^{6}$ Hz. We take $ \varepsilon _{ir}\sim 10$,
the value of $\varepsilon _{1}$ at the low frequency limit of
Fig.1(c).

\begin{figure}[tbp]
\centering\epsfig{file=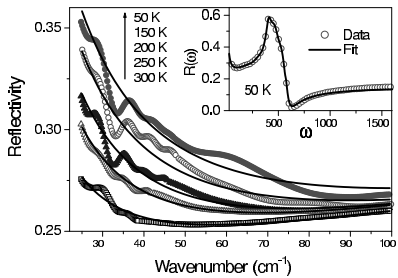,width=4.5cm,angle=-90}
\caption{T-dependence of the low frequency reflectivity. The solid lines are
the fitting curves using the Drude-Lorentz model, Eq.(1). Inset: the fit
over wide range of frequency for 50 K reflectivity.}
\label{fig:3}
\end{figure}

In the Maxwell-Wagner theory, high dielectric constant can arise in a
material with granular structure. Wu \textit{et al. } showed that LTNO
sample consists of the grains ($\sim $ $\mu m$ in size) separated by thin
boundary layer.\cite{wu} In such composite geometry, $\varepsilon (\omega )$
from the capacitance measurement is given as
\begin{eqnarray}
\frac{1}{\varepsilon (\omega )}=\frac{f_{g}}{\varepsilon
_{g}(\omega )}+ \frac{f_{b}}{\varepsilon _{b}(\omega )}
\end{eqnarray}
where ($\varepsilon _{g}(\omega ),f_{g}=1-f_{b}$) and
($\varepsilon _{b}(\omega ),$ $f_{b}$) represent the complex
dielectric constant and volume fraction of the grain and the
boundary, respectively. With $f_{b}\ll 1 $ the boundary
contribution is negligible to the infrared reflectivity and
R$(\omega)$ is dominated by the grain. From the effective medium
theory, the infrared result $\varepsilon (\omega )$ represents the
$\varepsilon _{g}(\omega )$.\cite{ema} As for $\varepsilon
_{b}(\omega )$, the boundary layer is thought to be highly
resistive: we measured the DC-resistivity $\rho $ of LTNO using
the four point probe method.\cite{resistivity} At all $T$, $ \rho$
was greater than the limit of the voltmeter, while $\rho\sim1$ ($
\Omega \cdot$ cm) is expected from the IR spectra (Table 1). This
suggests that the current of the Drude carrier is blocked by the
boundary. We have then $\varepsilon _{b}(\omega )=\varepsilon
_{ir}+i4\pi \sigma _{b}/\omega $ , $\sigma _{b}\ll \sigma _{g}$.
With these $\varepsilon _{g}(\omega )$ and $ \varepsilon
_{b}(\omega )$, we find that $\varepsilon(\omega )$ has the
Debye-type $\omega$-dependence, $\varepsilon (\omega )=\varepsilon
_{ir }+(\varepsilon _{o}-\varepsilon _{ir})/(1-i\omega /\Gamma )$.
Here the static dielectric constant $\varepsilon _{o}$ is given as
$\varepsilon _{ir}/f_{b}.$ If we take $f_{b}=10^{-3},$ the large
$\varepsilon _{0}=10^{4}$ is produced. On the other hand, the
crossover frequency $\Gamma $ is determined by the grain
conductivity. It is given approximately as the dc-value $\sigma
_{g}(o)\equiv\sigma_{g}$. From $\rho =1/\sigma_{g}$, we obtain
$\Gamma \sim 10^{10}$ Hz at 300 K. This value is far off from the
measured $\Gamma$ of Ref. 9, $\Gamma \sim 10^{7}$ Hz. Also,
$\sigma _{g}$ increases as $T$ decreases, which is the opposite of
the drastic decrease of the measured $\Gamma $. Moreover, for
$\omega >>\Gamma$, $\varepsilon (\omega )$ converges to
$\varepsilon _{ir}\sim 10$, which does not agree with the observed
value $\sim300$.\cite{wu} In this context, the two phase MW model
does not explain the dielectric constant of LTNO which suggests
that another ingredient should be involved.

\begin{table}[tbp]
\centering
\begin{tabular}[b]{cccccc}
\hline\hline
T(K) & 50 & 150 & 200 & 250 & 300 \\ \hline
$\omega_{p}$ (cm$^{-1}$) & 260.0 & 207.6 & 160.1 & 137.5 & 108.3 \\
$\gamma$ (cm$^{-1}$) & 287.7 & 208.7 & 144.1 & 123.6 & 98.8 \\
$\rho(\Omega \cdot$ cm) & 0.25 & 0.29 & 0.33 & 0.39 & 0.50 \\ \hline\hline
\end{tabular}
\label{Table 1} \caption{Fitting parameters of the Drude
conductivity: $\Omega_{p}$=plasma frequency,
$\protect\gamma$=scattering rate. The dc resistivity $\protect\rho
$ is calculated from $\Omega_{p}$ and $\protect\gamma$)(see text).
$\protect \varepsilon _{\infty }$=5.0 was used. The phonon
parameters are omitted for brevity}
\end{table}

Micro granular structure of oxide materials is often rather
complicated. In La-doped BaTiO3, for example, the outer region of
a grain is considered to have different electrical property from
the interior, and forms a third layer between the grain and the
grain boundary. West $et$ $al.$ used this three-layer picture to
explain the dielectric response of
La:BaTiO$_3$.\cite{west,sinclair2,saha} We consider this picture,
among others, as a possible explanation of our case: According to
a composition analysis of LTNO, Li ions are distributed within the
grain but Ti ions are segregated toward the boundary.\cite{wu} We
propose that they form a third layer between the grain and the
boundary. This layer will be more resistive than the boundary
layer: We assume that the boundary layer is bare NiO. It is known
that pristine NiO has residual Ni vacancies. The dilute vacancies
induce carriers (hole) which conduct through a thermally activated
hopping $ \sigma _{b}=\sigma _{o}e^{-U/kT}$ with $U\sim 0.1$
$eV$.\cite {niosemi,niosemi2} In the Ti-rich layer, the electrons
from Ti ions will compensate the holes from the Ni vacancies,
which makes the layer insulating ($\sigma =0$). This layer
contributes $f_{I}/\varepsilon _{I}(\omega )$ to Eq.(2) where
$f_{I}$( $<<1$) and $\varepsilon _{I}=\varepsilon _{ir}$ represent
the volume fraction and the dielectric constant, respectively.

The resulting $\varepsilon (\omega )$ is shown in Fig.4. In this
simulation, we used the parameters
$f_{b}=3\times10^{-2},f_{I}=10^{-3}$ and $U=0.3$ $eV$ . In the
upper panel, $\varepsilon (\omega )$ exhibits a two-step structure
where $\varepsilon (0)=10^{4}$ drops at $\omega =\Gamma _{L}$
($10^{4}$ Hz), followed by another drop at higher $\omega =\Gamma
_{H}$ ($10^{12}$ Hz). This behavior is understood from
$\varepsilon (\omega )$ of the three layers, $\varepsilon (\omega
)=\varepsilon _{ir}+i4\pi \sigma /\omega$ where
$\sigma_{g}\gg\sigma_{b}\gg\sigma_{I}=0$. When $\omega$ is high
enough, $ \varepsilon (\omega )$ of each layer becomes
$\varepsilon _{ir}$ and so does the total $\varepsilon (\omega )$.
This corresponds to the $\omega>\Gamma_{H} $ region. As $\omega $
decreases, the conducting term becomes comparable to $ \varepsilon
_{ir}$ for the grain layer, $\varepsilon_{ir}\sim 4\pi \sigma_{g}
/\omega$ while the other two layers are still insulating. This
situation corresponds to the two-component MW case of a conducting
grain and an effectively single layer of boundary with a volume
fraction of $ f_{b}+f_{I}$. Then $\varepsilon$ in this range is
given as $\varepsilon =\varepsilon _{ir}/(f_{b}+f_{I})$. The
crossover occurs when $\varepsilon _{ir}\sim\sigma _{g}/\omega $
which determines $\Gamma _{H}=$ $\sigma _{g}/\varepsilon _{ir}.$
Note that $\varepsilon $ is much higher than the infrared value
$\varepsilon _{ir}$, $\varepsilon \sim 30\times \varepsilon
_{ir}$. This result accounts for the puzzling discrepancy of the
dielectric constants from the infrared result and from the
microwave result. As $\omega$ is further decreased, the crossover
now occurs in the boundary layer when $ \varepsilon_{ir}\sim
4\pi\sigma_{b}/\omega$. Then a single effective grain is formed
and the MW composite has an insulating layer of $f_{I}$. Here $
\varepsilon =\varepsilon _{ir}/f_{I}$ and this crossover occurs at
$\Gamma _{L}=$ $\sigma _{b}/\varepsilon _{ir}.$

\begin{figure}[tbp]
\centering\epsfig{file=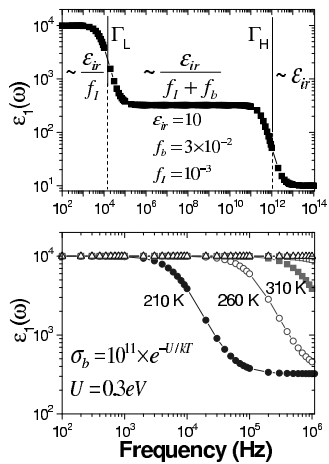,width=4.5cm,angle=0}
\caption{Dielectric constant simulated from the three phase MW model. (a)
Two step structure of $\protect\varepsilon(\protect\omega)$. $\Gamma_{L}$
and $\Gamma_{H}$ represent the the low and the high energy crossover
frequencies respectively. (b) Shift of $\Gamma_{L}$ with temperature.}
\label{fig:4}
\end{figure}

Unlike $\Gamma _{H}$, $\Gamma _{L}$ exhibits a large temperature
dependence. In the lower panel, we show $\varepsilon (\omega )$
for various $T$. As $T$ decreases, $\Gamma _{L}$ shifts rapidly
due to the T-dependence of $\sigma _{b}.$ This behavior closely
reproduces the experimental results. Also, the high dielectric
constant of LTNO is known to depend on the Li and Ti
concentration. If Ti becomes rich, in our picture, $f_{I}$ tends
to increase and $\varepsilon $ will decrease. On the other hand,
if Li content increases, the grain becomes Li-abundant. When
continued, even the boundary region will be overwhelmed by Li and
the Ti-layer will become thinner. Then $ f_{I}$ decreases and
$\varepsilon $ is enhanced, consistent with the observed behavior.
Although the model we have employed seems to explain the main
features of the dielectric phenomena of LTNO, it certainly needs
to be tested by further experiments. $\varepsilon (\omega )$
measurement  to observe the drop at  $\omega $= $\Gamma _{H}$ will
be useful. This requires an extension of the current experimental
window ( $\omega <10^{6}$ Hz) to higher fequency.  Also extension
of the reflectivity measurement toward the lower frequency,
perhaps using a terahertz spectroscopy and a resonant microwave
cavity method will be interesting.\ Further, a direct evidence of
the three layer structure using a scanning probe microscopy is
desired.

The Drude carrier is quite unconventional in that the plasma
frequency $ \omega _{p}$ increases as $T$ decreases while it is
conserved in normal metals. We note that NiO is a Mott-Hubbard
insulator and the doped LTNO can exhibit some unusual features of
the strong electron correlation origin. In a numerical study using
the dynamic mean field theory, the spectral weight of the Drude
conductivity ($\omega _{p}^{2}$) in a doped Hubbard system seems
to grow as $T$ decreases.\cite{dmft}Another unusual feature is the
increase of the scattering rate $\Gamma$ with decreasing $T$ which
is the opposite of the conventional behavior. This may suggest an
exotic scattering mechanism in LTNO which we do not understand at
this point.

In summary, we have studied the effect of Li and Ti doping in NiO
through the infrared spectroscopic measurement and considered the
implication on the high dielectric constant of LTNO. We find that
the doping brings about a Drude carrier which coexists with a
resistive grain boundary. This composite structure of electrically
different media support the Maxwell-Wagner mechanism as the origin
of the high $\varepsilon _{o}$. Our analysis showed that, however,
a two phase MW model does not account for the observed behaviors
of $\varepsilon (\omega )$ such as the frequency dependence. We
introduced a third layer, assumed to be insulating, and showed
that the three phase MW model can explain the high $\varepsilon $,
the $\omega -$ and $T-$ dependence. The large discrepancy between
the infrared $\varepsilon $ and the microwave $\varepsilon $ is
reconciled as well. For complete understanding of this material,
extension of the spectroscopic study into the intermediate region
between the infrared and the microwave frequencies, $ 10^{6}$ Hz
$< \omega$ $< 10^{11}$ Hz will be interesting.

This work was supported by the Korean Science and Engineering Foundation
(KOSEF)through Center for Strongly Correlated Materials Research (CSCMR).
One of us(EJC) acknowledges the financial support by the
KRF-2002-070-C00032. We thank useful discussions with H.-Y. Choi.

%
%
%

\bibliography{ltno}

\begin{thebibliography}{20}
\bibitem[{hig()}]{high}
\bibinfo{note}{We note that the reflectivity level is lower than the single
  crystal data in literature, probably due to the surface scattering and the
  low density of the polycrystal samples. We have performed a high temperature,
  high pressure annealing of NiO which resulted in an improved density. The
  reflectivity became higher close to that single crystals. However, the same
  treatment of LTNO caused a serious chemical dissociation unfortunately, and
  the sample lost the structural stability.}
\bibitem[{ema()}]{ema}
\bibinfo{note}{In the effective medium approximation, $\varepsilon$ is given
  from $f_b(\frac{\varepsilon _{b}-\varepsilon }{\varepsilon _{b}+2\varepsilon
  })+(1-f_b)(\frac{\varepsilon _{g}-\varepsilon }{\varepsilon _{g}+2\varepsilon
  })=0$. For $f_b\ll 1$, $\varepsilon $ converges to $\varepsilon _{g}$. This
  approxiamtion holds when the wavelength of incident radiation is much larger
  than the grain (and boundary) size, which is valid for our infrared
  measurement. Note that the infrared effective dielectric constant is
  different from that of the effective capacitance, Eq.(2).}
\bibitem[{res()}]{resistivity}
\bibinfo{note}{For the four probe resistivity measurement, we used a resistance
  bridge of PPMS (Physical Property Measurement System, Quantum Design). This
  instrument has a current source (I $>$ 10 nA) and a voltmeter (V $<$ 1 volt)
  which can measure resistance up to 10$^9 \Omega$.}
\end{thebibliography}

\end{document}